\documentclass[aps,prl,reprint,showpacs,floatfix,superscriptaddress]{revtex4-1}
\usepackage{color}

\usepackage[utf8]{inputenc}
\usepackage{amsmath,amssymb,amstext}
\usepackage{amsthm}
\usepackage{dsfont}
\usepackage{graphicx}

\def\sout{\bgroup\markoverwith
{\textcolor{red}{\rule[0.5ex]{2pt}{0.5pt}}}\ULon}
\def\be{\begin{equation}}
\def\ee{\end{equation}}
\def\bes{\begin{equation*}}
\def\ees{\end{equation*}}
\def\bea{\begin{eqnarray}}
\def\eea{\end{eqnarray}}
\def\beas{\begin{eqnarray*}}
\def\eeas{\end{eqnarray*}}
\def\bal#1\eal{\begin{align}#1\end{align}}
\def\bals#1\eals{\begin{align*}#1\end{align*}}
\newcommand{\bra}[1]{\langle #1|}
\newcommand{\ket}[1]{|#1\rangle}

\newcommand{\bk}[1]{\langle #1\rangle}

\renewcommand{\vec}{\vectorsym}

\renewcommand*{\vec}[1]{\boldsymbol{#1}}

\usepackage[normalem]{ulem}

\usepackage{braket}

\usepackage{pdfpages}
\usepackage{pgffor}

\makeatletter
\AtBeginDocument{\let\LS@rot\@undefined}
\makeatother

\begin{document}

\title{Molecular Impurities as a Realization of Anyons on the Two-Sphere}

\author{M. Brooks}
\email{morris.brooks@ist.ac.at}
\affiliation{IST Austria (Institute of Science and Technology Austria), Am Campus 1, 3400 Klosterneuburg, Austria}
\author{M. Lemeshko}
\affiliation{IST Austria (Institute of Science and Technology Austria), Am Campus 1, 3400 Klosterneuburg, Austria}
\author{D. Lundholm}
\affiliation{Uppsala University, Department of Mathematics - Box 480, SE-751 06 Uppsala, Sweden}
\author{E. Yakaboylu}
\email{enderalp.yakaboylu@ist.ac.at}
\affiliation{IST Austria (Institute of Science and Technology Austria), Am Campus 1, 3400 Klosterneuburg, Austria}

\date{\today}

\begin{abstract}

Studies on experimental realization of two-dimensional anyons in terms of quasiparticles have been restricted, so far, to only anyons on the plane. It is known, however, that the geometry and topology of space can have significant effects on quantum statistics for particles moving on it. Here, we have undertaken the first step towards realizing the emerging fractional statistics for particles restricted to move on the sphere, instead of on the plane. We show that such a model arises naturally in the context of quantum impurity problems. In particular, we demonstrate a setup in which the lowest-energy spectrum of two linear bosonic or fermionic molecules immersed in a quantum many-particle environment can coincide with the anyonic spectrum on the sphere. This paves the way towards experimental realization of anyons on the sphere using molecular impurities. Furthermore, since a change in the alignment of the molecules corresponds to the exchange of the particles on the sphere, such a realization reveals a novel type of exclusion principle for molecular impurities, which could also be of use as a powerful technique to measure the statistics parameter. Finally, our approach opens up a simple numerical route to investigate the spectra of many anyons on the sphere. Accordingly, we present the spectrum of two anyons on the sphere in the presence of a Dirac monopole field.

\end{abstract}

\maketitle

The study of quasiparticles with fractional statistics, called anyons, has been an active field of research in the past decades. This field has gained a lot of attention, due to the possible usage of these quasiparticles in quantum computation~\cite{K,lloyd2002quantum,freedman2003topological,Nayak_08}. In contrast to bosons and fermions, anyons acquire a phase $e^{i\pi \alpha}$ under the exchange of two particles, where the statistics parameter $\alpha$ is not necessarily an integer. The integer cases $\alpha =0 $ and $\alpha =1$  represent bosons and fermions, respectively. For non-integer $\alpha$, the transformation law $\Psi\rightarrow e^{i \pi\alpha} \Psi$ under the exchange of two particles, can only be realized by allowing the wave function $\Psi$ to be multivalued. The idea is that the multiple values keep book of the different possible ways the particles could ``braid" around each other. Due to the triviality of the braid group in $3+1$ dimensions, anyons are a purely low-dimensional phenomenon. 

Although anyons are predicted to be realized in the fractional quantum Hall effect (FQHE)~\cite{Tsui_82,Laughlin_83,Arovas_84,PhysRevLett.49.405,PhysRevLett.95.146802,
PhysRevLett.98.106803,PhysRevLett.61.2015,Lundholm_2016}, they have not yet been unambiguously detected in experiment. Indeed there has been a recent upsurge in interest concerning the realization of anyons in experimentally feasible systems~\cite{CooSim-15,ZhaSreGemJai-14,ZhaSreJai-15,MorTurPolWil-17,UmuMacComCar-18,CorDubLunRou-19}. For instance, it has been recently shown in Refs.~\cite{Yakaboylu_2018,Yakaboylu_anyon_2020} how these quasiparticles emerge from impurities in standard condensed matter systems. Nevertheless, all these works focus on the particles moving on the two-dimensional plane, i.e., on $\mathbb{R}^2$. Since the theory of anyons and their statistical behavior are strongly dependent on the geometry and topology of the underlying space, investigations on curved spaces reveal novel features of quantum statistics~\cite{Thouless_85,Einarsson_90,
einarsson1991fractional,Pithis_2015,
ouvry2019anyons,Tononi_2019,Tononi_2020,polychronakos2020two}. In particular, theoretical discussions for systems having various geometry and topology have widened our understanding of the FQHE~~\cite{Haldane_83,Laughlin_83}. 

In the present Letter, we explore the possibility of emerging fractional statistics for particles restricted to move on the sphere, $\mathbb{S}^2$, instead of on the plane. We show that such quasiparticles naturally arise from a system of molecular impurities exchanging angular momentum with a many-particle bath. In the regime of low energies, we identify the spectrum of this system with that of anyons. This does not only allow us to realize anyons on the sphere, but also to open up various numerical approaches to investigate the spectrum of $N$ anyons on the sphere. To illustrate this, we present the spectrum of two anyons on the sphere in the presence of a Dirac monopole field, extending the recent result of Ref.~\cite{ouvry2019anyons,polychronakos2020two}. Furthermore, the anyonic behavior of molecular impurities suggests that a novel type of exclusion principle holds, which concerns the alignment of the molecules, instead of the exchange of their actual position.

We start by considering a system of $N$ free anyons on the two-sphere. The Hamiltonian is given by the sum of the Laplacian of the $j$th particle on the sphere: $H_\text{0} = -\sum_{j=1}^N \nabla_j^2$, which acts on a multivalued wave function $\Psi$. By performing a singular gauge transformation, $\Psi\rightarrow e^{i\beta}\Psi$, one can get rid of the multivaluedness~\cite{Wu_84_anyon,mund1993hilbert,LO,O,R} and the free anyon Hamiltonian on the sphere~$H_\text{0}$ becomes equivalent to 
\be
\label{anyon_Ham_0}
H_{\mathrm{anyon}}=-\sum_{j =1}^N \left(\nabla_j-iA_j\right)^2 \,,
\ee
which now acts on single valued bosonic (fermionic) wave functions. Here anyons are depicted as bosons (fermions) interacting with the magnetic gauge field $A$, which explains that the calculation of the anyonic spectra is very hard~\cite{Lundholm_2017}. Note that $A=\nabla \beta$ is an almost pure gauge field, up to the singularities of $\beta$, where the particles meet, and it can be found as the variational solution of the Chern-Simons (CS) Lagrangian $L_{\mathbb{S}^2}=\sum_{j}\left(A \cdot \dot{q}_j+A_{0}\right)-(4\pi\alpha)^{-1}\int_{\mathbb{S}^2}\mathrm{d}\Omega\ A\wedge \mathrm{d}A$, where $q_j$ is the position of the nonrelativistic point particle coupled to the CS field, $A_0$ the time component of the gauge field, and $\wedge$ the wedge product. For anyons on the plane, one can always find a single magnetic potential $A$ as a solution. However, due to the non trivial homology of $\mathbb{S}^2$, the CS Lagrangian on the sphere can only be solved in two different stereographic coordinate charts: north and south patches, $A^N$ and $A^S$. As they should be a single object in the overlap patch, we require them to be gauge equivalent. This equivalence is given by the Dirac quantization condition (DQC) $ (N-1)\alpha\in  \mathbb{Z} $~\cite{LO,O}.

In what follows, in order to simplify our expressions, we represent the stereographic coordinates $(x,y)$ as a complex number, $z=x+iy$. In these coordinates, we define the gauge transformation $F=e^{i\beta}$, with $\beta(z_1,..,z_N)= - i \alpha \underset{j<k}{\sum}\log\left(\frac{z_j-z_k}{|z_j-z_k|}\right)$. The connections (gauge fields) are $A_{\bar{z}_j}= i D_{\bar{z}_j} \beta =- \frac{\alpha(1+|z_j|^2)}{2}\sum_{k\neq j}\left(\bar{z}_j-\bar{z}_k\right)^{-1}$ and $A_{z_j}= i D_{z_j} \beta = \frac{\alpha(1+|z_j|^2)}{2}\sum_{k\neq j}\left(z_j-z_k\right)^{-1}$, where we encode the contribution from the metric on $\mathbb{S}^2$ in the differential operators $D_{\bar{z}_j}=(1+|z_j|^2)\partial_{\bar{z}_j}$ and $D_{z_j}=(1+|z_j|^2)\partial_{z_j}$~\cite{CMO}. In the language of connections, $F$ represents the holonomy of $A$, and it is discontinuous along the lines which connect the particles with the north (south) pole, usually called the Dirac lines. Without loss of generality, we consider the north pole, which corresponds to the choice of $z_j = \cot(\theta_j/2)\exp(i \varphi_j)$, with spherical coordinates $\theta_j$ and $\varphi_j$. These lines represent the magnetic potential in the singular gauge, by assigning the particle an additional phase factor whenever it crosses them. The DQC makes sure that the Dirac lines are invisible, in the sense that one cannot distinguish between the theory where the lines run to the north pole and theories where they run to any other point. This means that our system is rotational invariant, up to gauge equivalences. 

The anyon Hamiltonian in our stereographic coordinate system is written as
\be
\label{anyon_ham}
H_{\mathrm{anyon}}=-\sum^N_{j=1}\left(D_{z_j}-\bar{z}_j-A_{z_j}\right)\left(D_{\bar{z}_j}-A_{\bar{z}_j}\right) \, .
\ee
Direct calculations to investigate the spectra of $H_{\mathrm{anyon}}$ turn out to be problematic, when the spectrum is calculated from the bosonic end. This is due to that the matrix elements of $A_{z_j} A_{\bar{z}_j}$ for certain bosonic states are singular, which is similar to the case of anyons on the plane~\cite{Yakaboylu_anyon_2020}. We can overcome this difficulty with the similarity transformation $H'_{\mathrm{anyon}} = e^{\alpha \sum_{j<k}\log|z_j-z_k|} H_{\mathrm{anyon}} e^{-\alpha \sum_{j<k}\log|z_j-z_k|}$. The advantage is that one of the two magnetic potentials vanishes in this pseudo-gauge and the Hamiltonian simplifies to
\be
\label{anyon_ham_weighted}
H'_{\mathrm{anyon}}=-\sum^N_{j=1}\left(D_{z_j}-\bar{z}_j-A'_{z_j}\right)D_{\bar{z}_j} \, ,
\ee
where the non-zero magnetic potential is $A'_{z_j} = 2 A_{z_j}$. Note that $H'_{\mathrm{anyon}}$ is self-adjoint in a weighted $L^2$ space. As we discuss below, while the first form of the anyon Hamiltonian~\eqref{anyon_ham} allows us to realize anyons in natural quantum impurity setups, the Hamiltonian~\eqref{anyon_ham_weighted} provides powerful numerical techniques to calculate the spectra of anyons on the sphere within the simplified impurity models.

We will now consider a general impurity problem of $N$ bosonic or fermionic impurities on $\mathbb{S}^2$ interacting with some Fock space $\mathcal{F}$. Within the Bogoliubov-Fr\"{o}hlich theory~\cite{frohlich1954electrons,bogolyubov1947theory,Pitaevskii2016}, the impurity Hamiltonian is
\bal
\label{gen_imp_ham}
& H_{\mathrm{imp}} = -\sum_{j=1}^N\left(D_{z_j}-\bar{z}_j\right)D_{\bar{z}_j}+\sum_{v}\omega_v b_v^\dagger b_v \\
\nonumber & + \sum_{v}\lambda_v(z_1,..,z_N)\left(e^{-i\beta_v(z_1,..,z_N)} b_v^\dagger+e^{i\beta_v(z_1,..,z_N)} b_v\right) \, ,
\eal
where $b_v^\dagger,b_v$ are the bosonic creation and annihilation operators in $\mathcal{F}$, $\omega_v$ is the energy of the mode $v$, and $\lambda_v(z_1,..,z_N)$ and $ \beta_v(z_1,..,z_N)$ describe the interaction of the impurities with the Fock space, depending on their coordinates $z_1,..,z_N$.  In the limit of $\omega_v\rightarrow \infty$ (the adiabatic limit), one can justify that the lowest spectrum of $H_{\mathrm{imp}}$ is described by the Born-Oppenheimer (BO) approximation; see Ref~\cite{Yakaboylu_anyon_2020} for an analysis of this assumption in the planar case. The projection of the Hamiltonian to the smaller Hilbert space manifests itself as a minimal coupling of the otherwise free particles with effective magnetic potentials $A_{z_1},..,A_{z_N}$ and a scalar potential $\Phi$. 

Following Ref~\cite{Yakaboylu_anyon_2020}, we first apply the transformation $S(z_1,..,z_N)=e^{-i\sum_v \beta_v b_v^\dagger b_v}$ to Eq.~\eqref{gen_imp_ham}, and then project the transformed Hamiltonian onto the coherent state $\ket{\phi (z_1,..,z_N) }=e^{-\sum_v \frac{\lambda_v}{\omega_v}(b^\dagger_v-b_v)}\ket{0}$. The emerging magnetic potential in complex coordinates is then given by
\begin{align}
\label{eq:imp}
A^{imp}_{z_j}= i\sum_v \left(\frac{\lambda_v}{\omega_v}\right)^2 D_{z_j}\beta_v \, .
\end{align}
Let us consider the specific choice  $\beta_v(z_1,..,z_N)= - i p_v \underset{j<k}{\sum}\log\left(\frac{z_j-z_k}{|z_j-z_k|}\right)$, which results in $A^{\mathrm{imp}}_{z_j}=\frac{\alpha(1+|z_j|^2)}{2}\sum_{k\neq j}\left( z_j-z_k \right)^{-1}$ with $\alpha(z_1,..,z_N)=\sum_v p_v \left(\frac{\lambda_v}{\omega_v}\right)^2$. We thus see that $A^{\mathrm{imp}}_{\bar{z}_j}$ is the sought CS gauge field and obeys the DQC if $\alpha(z_1,..,z_N)$ is a constant and satisfies $(N-1)\alpha\in \mathbb{Z}$. We emphasize, however, that for the values of $\alpha$ which do not satisfy the DQC, the impurity Hamiltonian~\eqref{gen_imp_ham} is still well-defined. The only difference for these values is that the theory is no longer fully rotational invariant, but, instead, it is invariant under rotation around the $z$ axis. In other words, the Dirac lines, which emerge together with the statistical gauge field, are not invisible~\cite{PRS} and they puncture the sphere. These features have drastic effects on the physical realization of anyons on the sphere in terms of quantum impurities, in comparison to emergent anyons on the plane studied in Ref.~\cite{Yakaboylu_anyon_2020}. 


\begin{figure}[t]
\centering
\includegraphics[width=\linewidth]{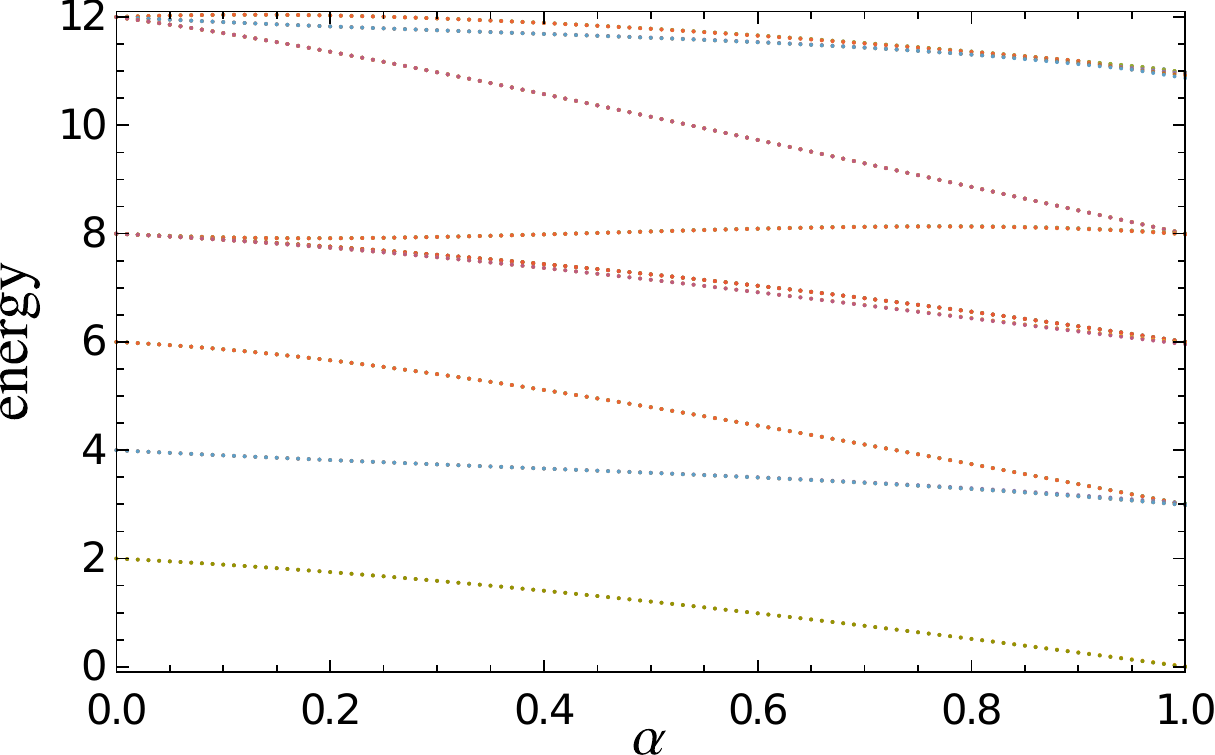}
\caption{Numerical computations of the energy of two anyons on the sphere in the presence of a Dirac monopole in terms of the relative statistics parameter, i.e., $\alpha = 0$ corresponds to fermions and $\alpha = 1$ to bosons. We set $2B = \alpha$ and consider spherical harmonics with the angular momentum up to $l_{\mathrm{max}}=8$ for the numerics. Compare Fig.~1 in Ref.~\cite{polychronakos2020two}.}
\label{fig_spec}
\end{figure}

In general, the impurity Hamiltonian~\eqref{gen_imp_ham} corresponds to \textit{interacting} anyons due the presence of the scalar potential $\Phi$. An impurity Hamiltonian whose lowest-energy spectrum is governed by the anyon Hamiltonian in the pseudo-gauge~\eqref{anyon_ham_weighted}, however, describes free anyons, as the scalar potential vanishes with $A_{\bar{z}} = 0$. Although such an impurity Hamiltonian is not Hermitian and may be harder to realize experimentally, considered as a toy model its non-Hermiticity is harmless for our purposes and it opens up simple numerical approaches to calculate the spectra of anyons on the sphere. 

Our numerical tools work for an arbitrary number of particles. Nevertheless, we will here study only the two-anyon case, since the computational effort strongly scales with the number of particles. Furthermore, we investigate impurities interacting with a Dirac monopole field $B$. This allows us to investigate the spectrum for all values of $\alpha$, as the DQC in the presence of a Dirac monopole field is $2 B - (N-1) \alpha \in \mathbb{Z}$~\cite{ouvry2019anyons,polychronakos2020two}. Accordingly, we consider the following simple model
\bal
\label{imp_ham_nonher}
& H'_{\mathrm{imp}}=H_B + \omega \left(b^\dagger b+\frac{\alpha}{p} \right) \\
\nonumber & +\sqrt{\frac{\alpha}{p}}\omega\left(e^{-p\log\left(z_1-z_2\right)} b^\dagger+e^{p\log\left(z_1-z_2\right)} b\right) \,,
\eal
where $H_B = H_0 +  \sum^2_{j=1} A^B_{z_j} D_{\bar{z}_j}$ describes the bosonic or fermionic particles interacting with the Dirac monopole field $B$ generated by the gauge field $A^B_{z_j} =2 B \bar{z}_j$, $p$ is an integer, and we subtracted the vacuum energy, $-\omega \alpha/p$, of the pure Fock space part of the Hamiltonian. 

One could calculate the lowest spectrum of $H'_{\mathrm{imp}}$ by diagonalizing the matrix $\bra{S (A) ; n} H'_{\mathrm{imp}}\ket{S' (A') ; n'}$, where $\ket{S (A)} = \ket{ Y_{l_1,m_1}\otimes_{S (A)} Y_{l_2,m_2}}$ are the impurity basis with $Y_{l,m}$ being the spherical harmonics, $\otimes_{S (A)}$ the (anti-)symmetric tensor product, and $\ket{n}$ the $n$-particle state in the Fock space. Instead of this direct diagonalization technique, we first diagonalize the Fock space part of the Hamiltonian with the displacement operator. The anyon Hamiltonian~\eqref{anyon_ham_weighted} in the presence of a Dirac monopole field, which emerges in the limit of $\omega \to \infty$, is, then, given by
\be
\label{anyon_ham_from_coh}
H'^{B}_{\mathrm{anyon}}=H_B+\frac{\alpha}{p}\left(e^{p\log\left(z_1-z_2\right)} H_{\mathrm{0}} e^{-p\log\left(z_1-z_2\right)}-H_{\mathrm{0}}\right) \, ,
\ee
see Supplemental Material for the derivation. We underline that a similar form of the Hamiltonian~\eqref{anyon_ham_from_coh} for anyons on the plane has been previously introduced in Ref.~\cite{Yakaboylu_anyon_2020}, where the second term of the right hand side was written in terms of composite bosons/fermions for an even integer $p$. Extending this approach we use here Bose-Fermi mixtures which enable us to set $p=1$. Within such a simple choice Eq.~\eqref{anyon_ham_from_coh} can be written as the following matrix equation
\be
\label{simple_calc}
E^B_{\mathrm{anyon}}=E_\text{bos} + 2 B \, W_S +\alpha\left(Z^{-1} E_{\mathrm{fer}} Z-E_{\mathrm{bos}}\right) \, ,
\ee
where the elements of the matrices are given by $E_{\mathrm{bos}}= \bra{S} H_0 \ket{S'}$, $E_{\mathrm{fer}}= \bra{A} H_0 \ket{A'}$, $W_S= \bra{S} \sum^2_{j=1} \bar{z}_j D_{\bar{z}_j} \ket{S'}$, and $Z^{-1} = \bra{S} z_1-z_2 \ket{A} $. As the latter two terms are straightforward to calculate numerically, and the matrix $Z$ can be obtained by taking the (pseudo)inverse of $Z^{-1}$, Eq.~\eqref{simple_calc} opens up a powerful route to calculate the anyonic spectrum. The spectrum from the fermionic end in terms of the relative statistics parameter can be calculated simply with the replacement of the basis $\ket{S (A)} \to \ket{A (S)}$ in Eq.~\eqref{simple_calc}.

As an example, we compute the eigenvalues for $\alpha$ ranging from $0$ to $1$. For an easier comparison with the result existing in Ref.~\cite{polychronakos2020two}, we calculate the 
spectrum from the fermionic end. The result presented in Fig.~\ref{fig_spec} is consistent with the one shown in Ref.~\cite{polychronakos2020two}, where the spectrum was calculated only for the subset of energy levels with unit total angular momentum.

\begin{figure}[t]
\centering
\includegraphics[width=\linewidth]{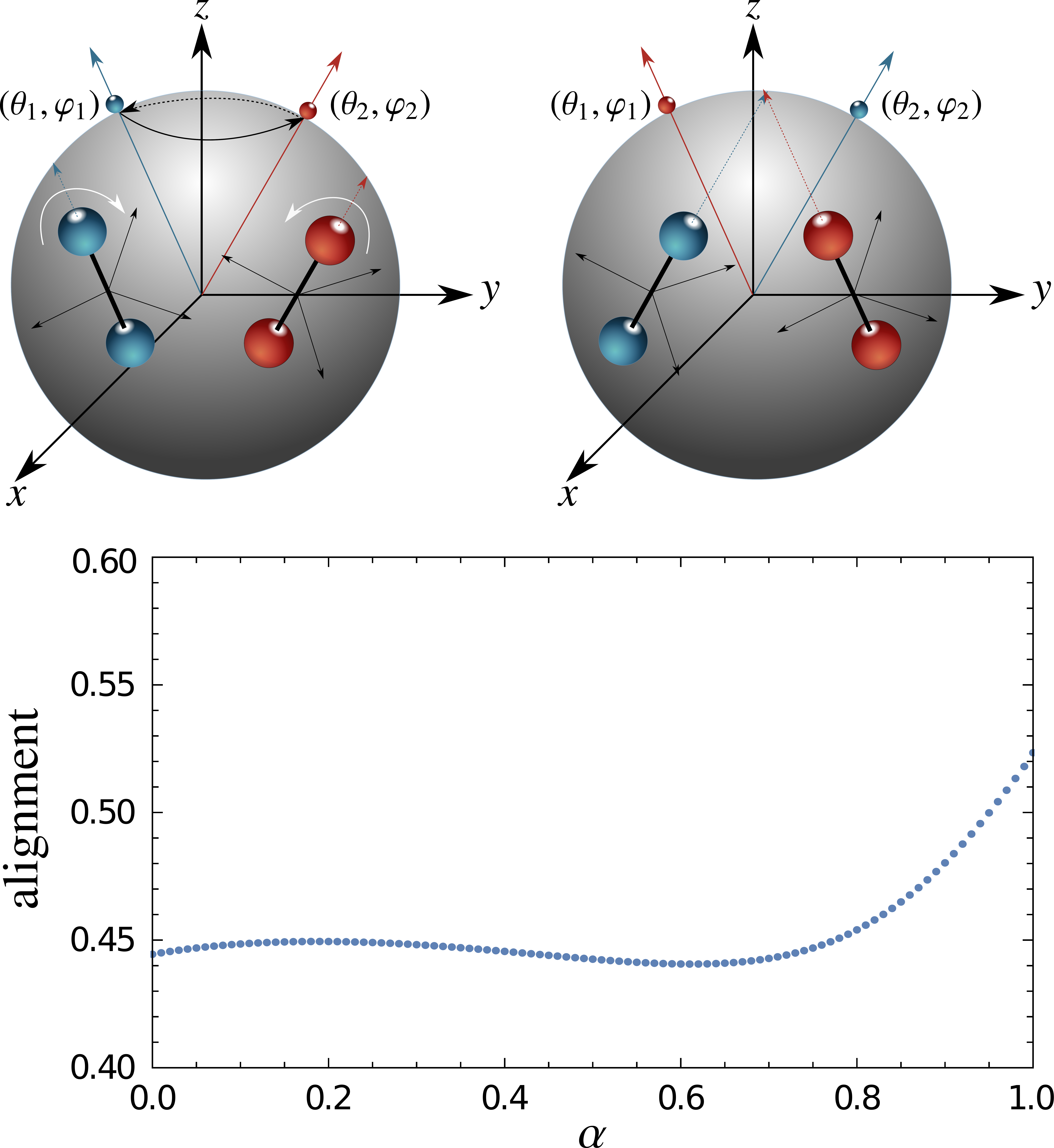}
\caption{(Top) Realization of anyons on the sphere in terms of linear molecules immersed in a quantum many-particle environment. A change in the alignment of the molecules (dumbbells), which is depicted by the white arrows, corresponds to the exchange of the particles on the sphere (dots), shown by the curvy black arrows. (Bottom) The alignment $\bk{\left( \cos \theta_1 - \cos \theta_2 \right)^2 }$ as a function of the absolute statistics parameter for the ground state. The curve follows the bosonic state $\ket{ Y_{0,0 }\otimes_{S} Y_{0,0}}$ at $\alpha=0$ to the fermionic state $\ket{ Y_{1,0}\otimes_{A} Y_{0,0}}$ at $\alpha=1$. We consider spherical harmonics with the angular momentum up to $l_{\mathrm{max}}=8$ for the numerics.}
\label{fig_conf}
\end{figure}

The general form of the impurity Hamiltonian~\eqref{gen_imp_ham} allow us also to physically realize anyons on the sphere in terms of quantum impurities. The kinetic energy of the particles on the sphere, which is given by the Laplacian, $-\left(D_{z_j}-\bar{z}_j \right)D_{\bar{z}_j}$, can be realized as the angular momentum operator $\vec{L}_j^2$. The latter can be considered as the Hamiltonian of linear molecules, which enables us to map rotation of molecules to motion of point particles on the sphere. Consequently, instead of point-like impurities, which have been considered for the planar case in Ref.~\cite{Yakaboylu_anyon_2020}, we consider here linear molecules and explore the angular momentum exchange with the bath. Such a realization exposes a novel correlation between molecular impurities. Specifically, the exchange of the particles on the sphere corresponds to a change in the alignment of the molecules, but not the exchange of the molecules themselves, see Fig.~\ref{fig_conf} (Top). Therefore, the emerging statistical interaction manifests itself in the alignment of molecules. 

To illustrate this in a transparent way, we consider the simple impurity Hamiltonian~\eqref{imp_ham_nonher} in the absence of the Dirac monopole. We investigate the alignment $\langle \left( \cos \theta_1 - \cos \theta_2 \right)^2 \rangle$ as a function of the statistics parameter for two molecules. In Fig.~\ref{fig_conf} (Bottom) we present the alignment for the ground state, which is obtained from Eq.~\eqref{simple_calc} for the case of $B=0$. We note that the Hamiltonian is still well-defined for the values of $\alpha$ which do not satisfy the DQC as we discussed before. Thus, the alignment of the molecules could be used as an experimental measure of the statistics parameter. Such a measurement can be performed, for instance, within the technique of laser-induced molecular alignment~\cite{Friedrich_95,lemeshko2013manipulation}. Further discussion of the alignment of molecules as a consequence of the statistical interaction will be the subject of future work.

A physical realization of the interaction between the molecules and a bath is also natural in the physics of impurities. Indeed, it was shown that the molecular impurities rotating in superfluid helium can be described within an impurity problem~\cite{Lemeshko_2015, PhysRevX.6.011012, Lemeshko_2016_book}. The resulting quasiparticle, which is called the angulon, represents a quantum impurity exchanging orbital angular momentum with a bath of quantum oscillators, and serves as a reliable model for the rotation of molecules in superfluids~\cite{lemeshko2016quasiparticle}. Therefore, we consider the following angulon Hamiltonian~\cite{Yakaboylu_pol,li2019intermolecular}
\bal
\label{hamilton:realization}
& H_\text{angulon}  = \sum^2_{j=1}\vec{L}_j^2 + V(q_1,q_2)+\sum_{k,l,m}\omega_{k,l,m} b_{k,l,m}^\dagger b_{k,l,m} \\
\nonumber & +\sum_{k,l,m}\lambda_{k,l,m}(q_1,q_2)\left(e^{-i \beta_{k,l,m}(q_1,q_2)}b^\dagger_{k,l,m}+ \text{H.c.} \right) \, ,
\eal
where $\hat{b}^\dagger_{k,l,m} $ and $\hat{b}_{k,l,m}$ are the bosonic creation and annihilation operators written in the spherical basis~\cite{Lemeshko_2015}, $q_i = (\theta_i , \varphi_i)$ are the angular coordinates representing the molecular rotation of the $i$-th molecule, $V$ is a confining potential, and $\text{H.c.}$ stands for Hermitian conjugate. Note that the coupling terms might depend on the intermolecular distance, as well. For heavy molecules the BO approximation can be justified with a gapped dispersion $\omega_{k,l,m}$. Furthermore, following our previous reasoning and Eq.~\eqref{eq:imp}, if the impurity-bath coupling satisfies the relation $ i\sum_{k,l,m} \left(\frac{\lambda_{k,l,m}}{\omega_{k,l,m}}\right)^2 D_{z_j}\beta_{k,l,m} = A_{z_j}$, the lowest-energy spectrum of the two linear molecules immersed in the bath coincide with the spectrum of two anyons on the sphere. In principle, such an interaction is feasible with the state-of-art techniques in the physics of superfluid helium as well as ultracold molecules.

In order to present a simple and intuitive realization, we first neglect the intermolecular distance. This enables us to define the interaction term simply as $\lambda_{k,l,m}(q_1,q_2) e^{-i \beta_{k,l,m}(q_1,q_2)} = u_{k,l} \sum_{j=1}^2 Y_{l,m}(q_j)$ with the impurity-bath coupling $u_{k,l}$. For a physical configuration, we consider molecular impurities in superfluid helium nanodroplets. The corresponding coupling captures the details of the molecule-helium interaction. For the form of the coupling and the relevant parameters we refer the reader to  Supplemental Material and Ref.~\cite{Cherepanov_2017,cherepanov2019far}, where the model has been used in order to describe angulon instabilities and oscillations observed in the experiment. Furthermore, the dispersion relation of superfluid helium allows us to achieve a gapped dispersion at the roton minimum~$\omega_r$~\cite{Lemeshko_2016_book}. Following the experimental realization proposed in Ref.~\cite{Yakaboylu_anyon_2020} for anyons on the plane,  we also couple the impurities to an additional constant magnetic field and rotate the whole system at the cyclotron frequency $\Omega$, which breaks time reversal symmetry so that anyons can emerge on the lowest-energy spectrum.

\begin{figure}
\centering
\includegraphics[width=\linewidth]{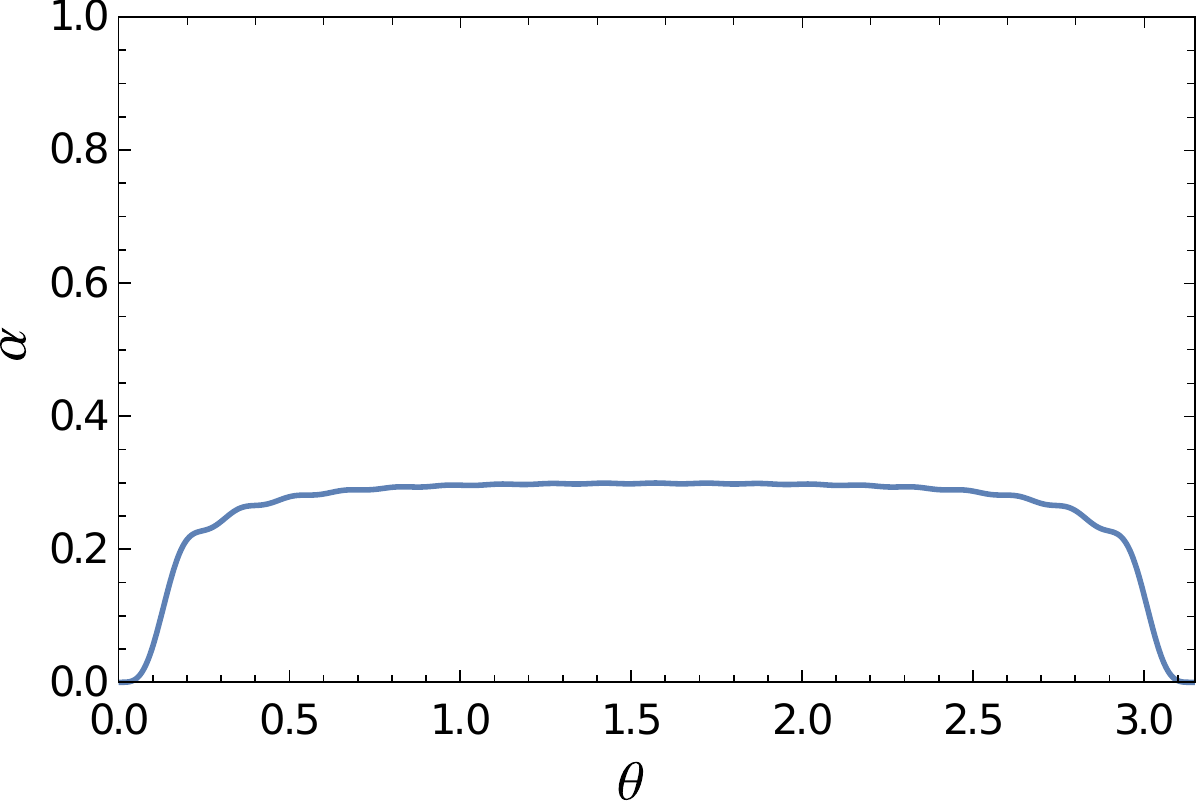}
\caption{The dependence of the statistics parameter $\alpha$ on the relative angle $\theta$. The computation is performed for the parameters modeling the molecule-helium interaction, given in Supplemental Material. The other parameters are $\omega_r =1$, $\Omega = 1.1$, and $l_\text{max} = 20$.}
\label{fig_real}
\end{figure}

A priori, the emerging statistics parameter $\alpha=\alpha(\theta)$ depends on the relative angle $\theta$ between the points $q_1$ and $q_2$. However, with a careful choice of the model parameters, $\alpha$ becomes approximately constant with the condition $\Omega \, l_\text{max} / \omega_r \gg1 $, see Supplemental Material. The condition imposes that the cyclotron frequency should be at the order of the roton minimum. This implies that molecular impurities should be subjected to a strong magnetic field at the order of $M \omega_r$ with $M$ being the mass of the molecules. The $\theta$ dependence of $\alpha$ is demonstrated in Fig.~\ref{fig_real}. In general, the statistics parameter does not satisfy the DQC. Therefore, the molecular impurities correspond to anyons interacting with the magnetic potential depicted by the Dirac lines, with broken rotational symmetry. We also note that with the additional confining potential, $V$, the particles are confined to one of the half spheres so that the statistics parameter becomes accessible to the experiment.

Thus, we see that a system of two linear molecules exchanging angular momentum with a many-particle bath can give rise to a system of quasiparticles with anyonic statistics, and can be realized by considering molecular impurities in superfluid helium droplets. It would be interesting to continue this approach and investigate, whether one can generalize the results above e.g. to non-Abelian Chern-Simons particles with the help of a higher order Born-Oppenheimer approximation.

\begin{acknowledgments}

We are grateful to A. Ghazaryan for valuable discussions and also thank the anonymous referees for comments. D.L. acknowledges financial support from the G\"oran Gustafsson Foundation (grant no.~1804) and LMU Munich. M.L. gratefully acknowledges financial support by the European Research Council (ERC) under the European Union's Horizon 2020 research and innovation programme (grant agreements No 801770).

\end{acknowledgments}

\bibliography{AnyonsBib}




\section*{Supplementary material}

M. Brooks,[1, *] M. Lemeshko,[1] D. Lundholm,[2] and E. Yakaboylu[1, †]

[1]*IST Austria (Institute of Science and Technology Austria),*
*Am Campus 1, 3400 Klosterneuburg, Austria*

[2]*Uppsala University, Department of Mathematics - Box 480, SE-751 06 Uppsala, Sweden*

Here we provide some further technical details for the Letter.

## I. THE NEW ANYON HAMILTONIAN

We start by considering Eq. (6) in the Letter

$$H'_{\text{imp}} = H_B + \omega \left( b^\dagger b + \frac{\alpha}{p} \right) + \sqrt{\frac{\alpha}{p}} \omega \left( e^{-p \log(z_1 - z_2)} b^\dagger + e^{p \log(z_1 - z_2)} b \right). \tag{1}$$

The Fock space part of the Hamiltonian can be diagonalized with the following displacement operator

$$T = \exp\left[ -\sqrt{\frac{\alpha}{p}} \left( e^{-p \log(z_1 - z_2)} b^\dagger - e^{p \log(z_1 - z_2)} b \right) \right]. \tag{2}$$

The transformed Hamiltonian is written as

$$T^{-1} H'_{\text{imp}} T = T^{-1} H_B T + \omega b^\dagger b. \tag{3}$$

In the limit of $\omega \to \infty$ the vacuum sector decouples from the rest of the spectrum, and the lowest energy levels are given by the anyon Hamiltonian in the pseudo gauge (see [20] for further details)

$$H'^B_{\text{anyon}} = \langle 0 | T^{-1} H'_{\text{imp}} T | 0 \rangle = \langle 0 | T^{-1} H_B T | 0 \rangle = \sum_{n=0}^{\infty} \frac{e^{-\alpha/p}}{n!} \left( \frac{\alpha}{p} \right)^n e^{np \log(z_1 - z_2)} H_B e^{-np \log(z_1 - z_2)}, \tag{4}$$

where we made use of the coherent state. After using the series expansion for $e^{-\alpha/p}$ and the double sum identity, Eq. (4) can be written as

$$H'^B_{\text{anyon}} = \sum_{n=0}^{\infty} \frac{1}{n!} \left( \frac{\alpha}{p} \right)^n K_n, \quad \text{with} \quad K_n = \sum_{j=0}^{n} \binom{n}{j} (-1)^{n-j} e^{jp \log(z_1 - z_2)} H_B e^{-jp \log(z_1 - z_2)}. \tag{5}$$

Using that $\log(z_1 - z_2)$ is a holomorphic function, i.e., $D_{\bar{z}} \log(z) = 0$, and $H_B = H_0 + \sum_{j=1}^{2} A^B_{z_j} D_{\bar{z}_j}$, $K_n = 0$ for all $n \geq 2$, the anyon Hamiltonian is further simplified to

$$H'^B_{\text{anyon}} = H_B + \frac{\alpha}{p} \left( e^{p \log(z_1 - z_2)} H_0 e^{-p \log(z_1 - z_2)} - H_0 \right). \tag{6}$$

---

* morris.brooks@ist.ac.at
† enderalp.yakaboylu@ist.ac.at

## II. EMERGING ANYONS FROM THE ANGULON HAMILTONIAN

Let us first consider the angulon Hamiltonian

$$H_{\text{angulon}} = -\sum_{j=1}^{2} \left(D_{z_j} - \bar{z}_j\right) D_{\bar{z}_j} + \sum_{k,l,m} \omega_{k,l} b^{\dagger}_{k,l,m} b_{k,l,m}$$
$$+ \sum_{j=1}^{2} \sum_{k,l,m} u_{k,l} \left(Y_{l,m}(\theta_j, \phi_j) b^{\dagger}_{k,l,m} + Y^{*}_{l,m}(\theta_j, \phi_j) b_{k,l,m}\right).$$

In the following, we want to derive the emerging magnetic fields as $\omega \to \infty$. Since the anyon gauge field only depends on the relative position of the particles to each other, we start our analysis by fixing the center of mass coordinate $q_{\text{center}} = (q_1 + q_2)/|q_1 + q_2|$. In order to do so, let $T = T(q_{\text{center}})$ be a rotation which maps the $q_{\text{center}}$ into the south pole and $T_{\mathcal{F}} = T_{\mathcal{F}}(q_{\text{center}})$ its promotion to a rotation on the whole Fock space. Furthermore, let $(\theta, \phi)$ be the spherical coordinates of the rotated point $\bar{q}_1 = T(q_{\text{center}}) q_1$. The transformed Hamilton operator $H'_{\text{angulon}} = T_{\mathcal{F}} H_{\text{angulon}} T_{\mathcal{F}}^{-1}$ then reads

$$H'_{\text{angulon}} = -\sum_{j=1}^{2} \left(D_{z_j} - \bar{z}_j - T_{\mathcal{F}} D_{z_j} T_{\mathcal{F}}^{-1}\right) \left(D_{\bar{z}_j} - T_{\mathcal{F}} D_{\bar{z}_j} T_{\mathcal{F}}^{-1}\right)$$
$$+ \sum_{k,l,m} \omega_{k,l} b^{\dagger}_{k,l,m} b_{k,l,m} + \sum_{k,l,m} \lambda_{k,l,m} \left(e^{im\phi} b^{\dagger}_{k,l,m} + e^{-im\phi} b_{k,l,m}\right),$$

with $\lambda_{k,l,m}(\theta) = u_{k,l} \left(Y_{l,m}(\theta, 0) + Y_{l,m}(\theta, \pi)\right)$. We define the function $\beta(q_1, q_2) = \phi$ as the polar angle of the rotated point $\bar{q}_1 = T(q_{\text{center}}) q_1$, which is indeed dependent on $q_1, q_2$. In the following, we want to think of $T_{\mathcal{F}} D_{z_j} T_{\mathcal{F}}^{-1}$ as a non-singular error term. To achieve this we have to confine the particles $q_1, q_2$ to a subregion $O \subset \mathbb{S}^2$ of the sphere, since $q_{\text{center}} = q_{\text{center}}(q_1, q_2)$ is singular for antipodal configurations $q_1 = -q_2$. A suitable choice would be an open set $O$ with $\bar{O} \subset \{q = (x,y,z) \in \mathbb{S}^2 : z < 0\}$. As a consequence, the map $(q_1, q_2) \mapsto T^{-1}(q_{\text{center}})$ is a smooth $SO(3)$ valued map, and therefore the complex derivative $TD_{\bar{z}_j}T^{-1}$ is a smooth $\mathbb{C} \otimes \mathfrak{so}(3)$ valued map. With this at hand, we can write its Fock space promotion as $T_{\mathcal{F}} D_{\bar{z}_j} T_{\mathcal{F}}^{-1} = f_j \Lambda_{\nu_j} + i g_j \Lambda_{\eta_j}$, where the $\Lambda_{\nu_j}, \Lambda_{\eta_j}$ are the angular momentum operators oriented along the $q_1, q_2$ dependent unit vectors $\nu_j = \nu_j(q_1, q_2) \in \mathbb{S}^2$, $\eta_j = \eta_j(q_1, q_2) \in \mathbb{S}^2$ and $f_j, g_j$ are smooth functions of $q_1, q_2$.

Similar to the planar case, we observe that the emerging theory for $H_{\text{angulon}}$ in its initial symmetric form leads to anyons with the statistics parameter $\alpha = 0$, i.e. non interacting bosons, since symmetry breaking is required. A more interesting theory emerges from the modified Hamilton

operator

$$H'_{\text{angulon},\Omega} = -\sum_{j=1}^{2} \left(D_{z_j} - \bar{z}_j - T_{\mathcal{F}} D_{z_j} T_{\mathcal{F}}^{-1}\right) \left(D_{\bar{z}_j} - T_{\mathcal{F}} D_{\bar{z}_j} T_{\mathcal{F}}^{-1}\right) + \Omega^2 V(q_1, q_2)$$
$$+ \sum_{k,l,m} (\omega_{k,l} + \Omega\, m)\, b^\dagger_{k,l,m} b_{k,l,m} + \sum_{k,l,m} \lambda_{k,l,m} \left(e^{im\phi} b^\dagger_{k,l,m} + e^{-im\phi} b_{k,l,m}\right),$$

where $V$ is a confining potential. Note that $H'_{\text{angulon},\Omega}$ is written in the $T_{\mathcal{F}}$ transformed picture. This means, that in the original picture the magnetic quantum number $m$ is oriented in the direction $q_{\text{center}}$. For anyons in the plane, a realization of $H'_{\text{angulon},\Omega}$ can be achieved by coupling $H_{\text{angulon}}$ to an additional constant magnetic field and rotating the whole system at the cyclotron frequency $\Omega$.

We now define $H_{emerg} = \langle \Psi | H'_{\text{angulon},\Omega} | \Psi \rangle$ as the 0-th order Born-Oppenheimer approximation, where $|\Psi\rangle$ is the vacuum sector corresponding to $H'_{\text{angulon},\Omega}$. Note that we can write $|\Psi\rangle = S |\Psi_0\rangle$, with

$$S = e^{i \sum_{k,l,m} m\phi\, b^\dagger_{k,l,m} b_{k,l,m}},$$
$$|\Psi_0\rangle = e^{-\sum_{k,l,m} \frac{\lambda_{k,l,m}}{\omega_{k,l} + \Omega\, m} \left(b^\dagger_{k,l,m} - b_{k,l,m}\right)} |0\rangle.$$

Let us abbreviate the operator $X_{z_j} = -i \sum_{k,l,m} m \left(D_{z_j} \phi\right) b^\dagger_{k,l,m} b_{k,l,m} + S T_{\mathcal{F}} D_{z_j} T_{\mathcal{F}}^{-1} S^{-1}$. A straightforward computation exhibits

$$S H'_{\text{angulon},\Omega} S^{-1} = -\sum_{j=1}^{2} \left(D_{z_j} - \bar{z}_j - X_{z_j}\right) \left(D_{\bar{z}_j} - X_{\bar{z}_j}\right) + \Omega^2 V$$
$$+ \sum_{k,l,m} (\omega_{k,l} + \Omega\, m) \left(b^\dagger_{k,l,m} + \frac{\lambda_{k,l,m}}{\omega_{k,l} + \Omega\, m}\right) \left(b_{k,l,m} + \frac{\lambda_{k,l,m}}{\omega_{k,l} + \Omega\, m}\right) - \sum_{k,l,m} \frac{\lambda^2_{k,l,m}}{\omega_{k,l} + \Omega\, m}.$$

It is an important observation that $\lambda_{k,l,m}/(\omega_{k,l} + \Omega\, m) \in \mathbb{R}$. As a consequence, we obtain $\langle \Psi_0 | D_{z_j} | \Psi_0 \rangle = 0$. With this at hand, we can express the 0-th order Born-Oppenheimer approximation as

$$H_{emerg} = \langle \Psi_0 | S H'_{\text{angulon},\Omega} S^{-1} | \Psi_0 \rangle = -\sum_{j=1}^{n} \left(D_{z_j} - \bar{z}_j - \tilde{A}_{z_j}\right) \left(D_{\bar{z}_j} - \tilde{A}_{\bar{z}_j}\right) + \Phi$$
$$- \sum_{j=1}^{n} \left(\delta_{z_j} D_{\bar{z}_j} + \delta_{\bar{z}_j} D_{z_j}\right),$$

where $\Phi$ is a scalar potential, $\delta_{z_j}$ an error contribution which can be neglected in the limit $\Omega \to \infty$ (see below), and where the emerging gauge field $\tilde{A}_{\bar{z}_j}$ is given by

$$\tilde{A}_{\bar{z}_j} = \alpha(\theta)\, D_{\bar{z}_j} \phi,$$

with $\alpha(\theta) = \sum_{k,l,m} m \left(\frac{\lambda_{k,l,m}(\theta)}{w_{k,l} + \Omega\, m}\right)^2$. We can explicitly write down the additional emerging quantities $\Phi$ and $\delta_{z_j}$ as

$$\delta_{\bar{z}_j} = \langle\Psi|T_\mathcal{F} D_{\bar{z}_j} T_\mathcal{F}^{-1}|\Psi\rangle = f_j \langle\Psi|\Lambda_{\nu_j}|\Psi\rangle + ig_j \langle\Psi|\Lambda_{\eta_j}|\Psi\rangle,$$

$$\Phi = \sum_{j=1}^{2} \left(\tilde{A}_{z_j}\tilde{A}_{\bar{z}_j} - D_{z_j}\tilde{A}_{\bar{z}_j} + \langle\Psi_0|X_{z_j}D_{\bar{z}_j} + D_{z_j}X_{\bar{z}_j} - D_{z_j}D_{\bar{z}_j} - X_{z_j}X_{\bar{z}_j}|\Psi_0\rangle\right)$$

$$- \sum_{k,l,m} \frac{\lambda_{k,l,m}^2}{\omega_{k,l} + \Omega\, m} + \Omega^2 V.$$

In order to neglect the error contribution $\delta_{\bar{z}_j}$, it is important to observe that $\delta_{\bar{z}_j}$, in contrast to $\tilde{A}_{\bar{z}_j}$, is a non-singular quantity. In the limit $\Omega \to \infty$, we expect $D_{\bar{z}_j}$ to scale like $\Omega$ and therefore the kinetic energy scales like $\Omega^2$ while $\delta_{z_j} D_{\bar{z}_j} + \delta_{\bar{z}_j} D_{z_j}$ only scales like $\Omega$. Therefore, we are going to neglect the error contribution (compare Ref. [20]).

Note that the statistics parameter $\alpha$ is a priori a $\theta$-dependent function

$$\alpha(\theta) = \sum_{k,l,m} m \left(\frac{\lambda_{k,l,m}(\theta)}{w_{k,l} + \Omega\, m}\right)^2$$

$$= \sum_{k,l,m} m \left(\frac{u_{k,l}\,(Y_{l,m}(\theta,0) + Y_{l,m}(\theta,\pi))}{w_{k,l} + \Omega\, m}\right)^2.$$

However, by choosing the interaction coefficients carefully, it is possible to make $\alpha$ approximately constant, which is discussed below.

### III. THE MODEL COUPLING FOR THE MOLECULE-HELIUM INTERACTION

The original derivation of the angulon Hamiltonian is based on the description of an ultracold molecule coupled to a weakly-interacting BEC within the Bogoliubov approximation and transformation. Nevertheless, the angulon Hamiltonian can be used to study any quantum system whose angular momentum is coupled to a bath of quantum oscillators. Indeed, it was shown in Refs. [47,50,51] that the angulon Hamiltonian can very well describe the molecular rotation in helium droplets with the following coupling

$$u_{k,l} = \sqrt{\frac{8n_0 k^2 \epsilon(k)}{\omega(k)(2\lambda + 1)}} \int dr\, r^2 f_l(k) j_l(kr), \qquad (7)$$

which models the molecule-helium interaction. Here $n_0$ is the helium density, $\epsilon(k) = k^2/2$ the kinetic energy of helium atoms in the atomic units, and $\omega(k)$ the dispersion relation of superfluid helium nanodroplets. In order to achieve a gapped dispersion we consider the dispersion at the

roton minimum, $\omega_r$ [46]. Furthermore, $j_l(kr)$ is the spherical Bessel function and $f_l(k)$ represent the components of the spherical harmonics expansion of the molecule-helium potentials. The latter can be further modeled with effective potentials characterized by the Gaussian form-factors $f_l(k) = v_l(2\pi)^{-3/2} e^{-r^2/\rho_l^2}$ such that the magnitude $v_l$ and range $\rho_l$ reproduce known properties of the molecule-helium interaction.

The statistics parameter for molecular impurities immersed in helium droplets, then, can be written as

$$\alpha(\theta) = \frac{n_0}{32\pi^{3/2}\omega_r} \sum_{l,m} m\, N_l \left( \frac{Y_{l,m}(\theta,0) + Y_{l,m}(\theta,\pi)}{\omega_r + \Omega\, m} \right)^2, \qquad (8)$$

where $N_l = v_l^2 \rho_l (3 + 4l(l+1))/(2l+1)$. We observe that if $\Omega\, l_{\max}/\omega_r \gg 1$ and the factor $N_l$ scales as $(2l+1)^{-2}$, then the statistics parameter becomes approximately constant for a large range of $\theta$, when we consider many components of the spherical harmonics expansion of molecule-helium potentials, i.e., $l_{\max} \gg 1$. The latter condition is already consistent with realistic molecule-atom potentials, see for instance Ref.[51]. Accordingly, we set $v_l = v_0/\sqrt{2l+1}$ and $\rho_l = \rho_0(3+4l(l+1))^{-1}$ with the following simple parameters $v_0 = 1$ and $\rho_0 = 1$. We further choose $\omega_r = 1$, $\Omega = 1.1$, and $l_{\max} = 20$. Finally, we consider the density as $n_0 = e^7$, which reflects the high density of states of helium droplets near the rotor minimum. The result is demonstrated in Fig. 3 in the Letter.

We note that for constant $\alpha$, the emerging magnetic field $\alpha \tilde{A}_{\bar{z}_j}$ is almost pure gauge, up to the singularities of $\tilde{A}_{\bar{z}_j}$. In case that $\alpha$ satisfies the Dirac quantization condition $\alpha \tilde{A}_{\bar{z}_j}$ is even gauge equivalent to the usual anyon field $\alpha A_{\bar{z}_j}$. One can verify this, by passing to the singular gauge via the transformation $e^{i\alpha\phi}$, which has no line of discontinuity as long as $\alpha \in \mathbb{Z}$ and the right behavior under the exchange of $z_1$ with $z_2$.

Thus, we have seen that in the limit $\Omega \to \infty$ the 0-th order Born-Oppenheimer approximation of the Hamiltonian $H'_{\text{angulon},\Omega}$ corresponds to anyons coupled to a scalar potential $\Phi$, given that the parameters are chosen suitable as described above and in Fig. 3 in the Letter. Note that for large $\Omega$, the additional confining potential $\Omega^2\, V$ also justifies our assumption that the particles are confined to one of the half spheres.



\end{document}